\documentclass[aps,showpacs,preprintnumbers,amsmath, amssymb]{revtex4}

\usepackage{float}
\usepackage{graphics,epsfig}
\usepackage{graphicx}
\usepackage{dcolumn}
\usepackage{bm}

\begin{document}
\baselineskip=0.8 cm
\title{{\bf Holographic p-wave superconductor models with Weyl corrections}}

\author{Lu Zhang$^{1,2}$, Qiyuan Pan$^{1,2,3}$\footnote{panqiyuan@126.com} and Jiliang Jing$^{1,2}$\footnote{jljing@hunnu.edu.cn}}
\affiliation{$^{1}$Institute of Physics and Department of Physics,
Hunan Normal University, Changsha, Hunan 410081, China}
\affiliation{$^{2}$ Key Laboratory of Low Dimensional Quantum
Structures and Quantum Control of Ministry of Education, Hunan
Normal University, Changsha, Hunan 410081, China}
\affiliation{$^{3}$ Instituto de F\'{\i}sica, Universidade de
S\~{a}o Paulo, CP 66318, S\~{a}o Paulo 05315-970, Brazil}

\vspace*{0.2cm}
\begin{abstract}
\baselineskip=0.6 cm
\begin{center}
{\bf Abstract}
\end{center}

We study the effect of the Weyl corrections on the holographic
p-wave dual models in the backgrounds of AdS soliton and AdS black
hole via a Maxwell complex vector field model by using the numerical
and analytical methods. We find that, in the soliton background, the
Weyl corrections do not influence the properties of the holographic
p-wave insulator/superconductor phase transition, which is different
from that of the Yang-Mills theory. However, in the black hole
background, we observe that similar to the Weyl correction effects
in the Yang-Mills theory, the higher Weyl corrections make it easier
for the p-wave metal/superconductor phase transition to be
triggered, which shows that these two p-wave models with Weyl
corrections share some similar features for the condensation of the
vector operator.

\end{abstract}
\maketitle
\newpage
\vspace*{0.2cm}

\section{Introduction}

As a brilliant concept, the anti-de Sitter/conformal field theory
(AdS/CFT) correspondence conjectures a duality between strongly
coupled quantum field theories and weakly coupled gravity theories
\cite{Maldacena}, which has become a powerful tool to study the
condensed matter systems. It was shown that a gravitational model of
hairy black holes \cite{GubserPRD78}, where the Abelian symmetry of
Higgs is spontaneously broken below some critical temperature, can
be used to model high $T_{c}$ superconductor \cite{HartnollPRL101}.
Interestingly, the properties of a ($2+1$)-dimensional
superconductor can indeed be reproduced in the ($3+1$)-dimensional
holographic dual model in the background of AdS black hole
\cite{HartnollJHEP12}. Extended the investigation to the bulk AdS
soliton background, it is found that when the chemical potential is
sufficiently large beyond a critical value $\mu_{c}$, the soliton
becomes unstable to form scalar hair and a second order phase
transition can happen, which can be used to describe the transition
between the insulator and superconductor
\cite{Nishioka-Ryu-Takayanagi}. In recent years, the so-called
holographic superconductor models have attracted a lot of attention;
for reviews, see Refs. \cite{S.A. HartnollRev,C.P. HerzogRev,G.T.
HorowitzRev} and the references therein.

In general, the studies on the holographic superconductors focus on
the Einstein-Maxwell theory coupled to a charged scalar field. In
order to understand the influences of the $1/N$ or $1/\lambda$
($\lambda$ is the 't Hooft coupling) corrections on the holographic
dual models, it is interesting to consider the curvature correction
to the gravity \cite{Gregory,Pan-Wang} and the higher derivative
correction related to the gauge field \cite{JS2010}. Recently, an
s-wave holographic superconductor model with Weyl corrections has
been introduced in order to explore the effects beyond the large $N$
limit on the superconductor \cite{WuCKW}. It was observed that,
unlike the effect of the higher curvature corrections
\cite{Gregory,Pan-Wang}, the higher Weyl corrections make it easier
for the condensation to form. Then, introducing an $SU(2)$
Yang-Mills action with Weyl corrections into the bulk, Momeni
\emph{et al.} studied the p-wave holographic superconductor with
Weyl corrections and found that the effect of Weyl corrections on
the condensation is similar to that of the s-wave model
\cite{MomeniSL}. Considering the holographic
insulator/superconductor phase transition model with Weyl
corrections to the usual Maxwell field in the probe limit, we found
that the higher Weyl corrections make the insulator/superconductor
phase transition harder to occur in p-wave model but will not affect
the properties of the insulator/superconductor phase transition in
s-wave case \cite{ZPJ2012}. Holographic superconductor models with
Weyl corrections can also be found, for example, in Refs.
\cite{MaCW,WeylMS,WeylRoychowdhury,WeylMSM,WeylMMR,ZhangLiZhao,WeylMRM}.

More recently, Cai \emph{et al.} constructed a new p-wave
holographic superconductor model by introducing a charged vector
field into an Einstein-Maxwell theory with a negative cosmological
constant \cite{CaiPWave-1}. In the probe limit, they obtained a
critical temperature at which the system undergoes a second order
phase transition and observed that an applied magnetic field can
induce the condensate even without the charge density. When taking
the backreaction into account, a rich phase structure: zeroth order,
first order and second order phase transitions in this p-wave model
has been found \cite{CaiPWave-2,CaiPWave-3}. Using a
five-dimensional AdS soliton background coupled to such a Maxwell
complex vector field, the authors of \cite{CaiPWave-4} reconstructed
the holographic p-wave insulator/superconductor phase transition
model in the probe limit and showed that the
Einstein-Maxwell-complex vector field model is a generalization of
the $SU(2)$ model with a general mass and gyromagnetic ratio. In
Ref. \cite{CaiPWave-5}, the complete phase diagrams of this new
p-wave model has been discussed by considering both the soliton and
black hole backgrounds. Other generalized investigations based on
this new p-wave model can be found, for example, in Refs.
\cite{NPWave-1,NPWave-2,NPWave-3,NPWave-4,NPWave-5}.

Considering that the increasing interest in study of the Maxwell
complex vector field model, in this work we will consider the new
p-wave holographic dual models with Weyl corrections to the usual
Maxwell field via the action
\begin{eqnarray}\label{NewPWction}
S=\frac{1}{16\pi G}\int
d^{5}x\sqrt{-g}\left[R+\frac{12}{L^2}-\frac{1}{4}\left(F_{\mu\nu}F^{\mu\nu}-4\gamma
C^{\mu\nu\rho\sigma}F_{\mu\nu}F_{\rho\sigma}\right)-\frac{1}{2}\rho_{\mu\nu}^{\dag}\rho^{\mu\nu}-m^2\rho_{\mu}^{\dag}\rho^{\mu}+i
q\gamma_0\rho_{\mu}\rho_{\nu}^{\dag}F^{\mu\nu} \right],
\end{eqnarray}
where $G$ is the gravitational constant in the bulk, $\gamma$ is the
Weyl coupling parameter which satisfies $-L^{2}/16<\gamma<L^{2}/24$
\cite{Ritz-Ward}, and $L$ is the AdS radius which will be chosen to
be unity. $m$ and $q$ represent the mass and charge of the vector
field $\rho_\mu$, respectively. The strength of $U(1)$ field $A_\mu$
is $F_{\mu\nu}=\nabla_{\mu}A_{\nu}-\nabla_{\nu}A_{\mu}$ and the
tensor $\rho_{\mu\nu}$ is defined by
$\rho_{\mu\nu}=D_\mu\rho_\nu-D_\nu\rho_\mu$ with the covariant
derivative $D_\mu=\nabla_\mu-iqA_\mu$. The parameter $\gamma_0$,
which describes the interaction between the vector field $\rho_\mu$
and the gauge field $A_\mu$, will not play any role because we will
consider the case without external magnetic field. Since the Weyl
corrections do have effects on the metal/superconductor
\cite{MomeniSL} and insulator/superconductor \cite{ZPJ2012} phase
transitions for the holographic p-wave dual models via the
Yang-Mills theory, we try to discuss the effect of the Weyl
corrections on this new p-wave holographic dual models, and want to
know the difference between these two p-wave models. In order to
extract the main physics, we will concentrate on the probe limit to
avoid the complex computation.

The structure of this work is as follows. In Sec. II we will
investigate the p-wave insulator/superconductor phase transition
with Weyl corrections of the Maxwell complex vector field which has
not been constructed as far as we know, and compare it with that of
the Yang-Mills theory. In Sec. III we extend the discussion to the
metal/superconductor case. We will conclude in the last section with
our main results.

\section{p-wave superconductor models with Weyl corrections in AdS soliton}

In Ref. \cite{ZPJ2012}, we considered an $SU(2)$ Yang-Mills action
with Weyl corrections in the bulk theory to construct the
holographic p-wave insulator/superconductor phase transition with
Weyl corrections and found that the higher corrections make the
phase transition harder to occur. Now we will study the effect of
the Weyl corrections on the new p-wave insulator/superconductor
phase transition via the Maxwell complex vector field model
(\ref{NewPWction}).

\subsection{Numerical investigation of holographic insulator/superconductor phase transition}

In order to study the superconducting phase dual with Weyl
corrections to the AdS soliton configuration in the probe limit, we
start with the five-dimensional Schwarzschild-AdS soliton in the
form
\begin{eqnarray}\label{SchSoliton}
ds^2=-r^2dt^2+\frac{dr^2}{f\left(r\right)}+f\left(r\right)d\varphi^2+r^2(dx^2+dy^2),
\end{eqnarray}
where $f(r)=r^2(1-r_{s}^{4}/r^{4})$ with the tip of the soliton
$r_{s}$ which is a conical singularity in this solution. By imposing
a period $\beta=\pi/r_{s}$ for the coordinate $\varphi$, we can
remove the singularity. For the considered solution
(\ref{SchSoliton}), the nonzero components of the Weyl tensor
$C_{\mu\nu\rho\sigma}$ are
\begin{eqnarray}
\begin{aligned}
&C_{0i0j}=-r_s^4\delta_{ij},\quad C_{0r0r}=\frac{r_s^4}{r^4-r_s^4} ,\quad C_{0\varphi0\varphi}=r_s^4\left(1-\frac{r_s^4}{r^4}\right),\quad C_{r\varphi r\varphi}=\frac{3r_s^4}{r^4},\\
&C_{irjr}=-\frac{r_s^4}{r^4-r_s^4}\delta_{ij},\quad C_{i\varphi j\varphi}=-r_s^4\left(1-\frac{r_s^4}{r^4}\right)\delta_{ij},\quad C_{ijkl}=r_s^4\delta_{ik}\delta_{jl},
\end{aligned}
\end{eqnarray}
with $i,~j,~k,~l=x$ or $y$.

Just as in Refs. \cite{CaiPWave-1,CaiPWave-2}, we assume the
condensate to pick out the $x$ direction as special and take the
following ansatz
\begin{eqnarray}\label{Ansatz}
\rho_{\nu}dx^{\nu}=\rho_{x}(r)dx\,,\hspace{0.5cm}A_{\nu}dx^{\nu}=\phi(r)dt,
\end{eqnarray}
where we can set $\rho_{x}$ to be real by using the $U(1)$ gauge
symmetry. Thus, we can obtain the equations of motion from the
action (\ref{NewPWction}) for the vector hair $\rho_{x}$ and gauge
field $\phi$
\begin{eqnarray}
\rho_{x}^{\prime\prime}+\left(\frac{1}{r}
+\frac{f^\prime}{f}\right)\rho_{x}^\prime +\left(\frac{q^2
\phi^2}{r^2f}-\frac{m^2}{f}\right)\rho_{x}=0\,, \label{SLRhoxEoM}
\end{eqnarray}
\begin{eqnarray}
\left(1+\frac{8\gamma
r_s^4}{r^4}\right)\phi^{\prime\prime}+\left[\frac{1}{r}\left(1-\frac{24\gamma
r_s^4}{r^4}\right)+\frac{f^\prime}{f}\left(1+\frac{8\gamma
r_s^4}{r^4}\right)\right]
\phi^\prime-\frac{2q^2\rho_x^2}{r^2f}\phi=0, \label{SLPhiEoM}
\end{eqnarray}
where the prime denotes the derivative with respect to $r$.

Using the shooting method \cite{HartnollPRL101}, we can solve
numerically the equations of motion (\ref{SLRhoxEoM}) and
(\ref{SLPhiEoM}) by doing integration from the tip out to the
infinity. At the tip $r=r_{s}$, the appropriate boundary conditions
for $\rho_x(r)$ and $\phi(r)$ are
\begin{eqnarray}\label{SLBoundary}
&&\rho_x=\tilde{\rho_x}_{0}+\tilde{\rho_x}_{1}(r-r_{s})+\tilde{\rho_x}_{2}(r-r_{s})^{2}+\cdots\,, \nonumber \\
&&\phi=\tilde{\phi}_{0}+\tilde{\phi}_{1}(r-r_{s})+\tilde{\phi}_{2}(r-r_{s})^{2}+\cdots\,,
\end{eqnarray}
where $\tilde{\rho_x}_{i}$ and $\tilde{\phi}_{i}$ ($i=0,1,2,\cdots$)
are the integration constants, and the Neumann-like boundary
conditions to render the physical quantities finite have been
imposed \cite{Nishioka-Ryu-Takayanagi}. It should be noted that
there is a constant nonzero gauge field $\phi(r_{s})$ at $r=r_{s}$,
which is in strong contrast to that of the AdS black hole where
$\phi(r_{+})=0$ at the horizon
\cite{HartnollPRL101,Nishioka-Ryu-Takayanagi}. At the asymptotic AdS
boundary $r\rightarrow\infty$, we have the boundary conditions
\begin{eqnarray}
\rho_{x}=\frac{\rho_{x-}}{r^{\Delta_{-}}}+\frac{\rho_{x+}}{r^{\Delta_{+}}}\,,\hspace{0.5cm}
\phi=\mu-\frac{\rho}{r^{2}}\,, \label{SLinfinity}
\end{eqnarray}
with the characteristic exponent $\Delta_\pm=1\pm\sqrt{1+m^{2}}$.
According to the AdS/CFT correspondence, $\mu$, $\rho$, $\rho_{x-}$
and $\rho_{x+}$ are interpreted as the chemical potential, the
charge density, the source and the vacuum expectation value of the
vector operator $J_{x}$ in the dual field theory respectively. In
this work, we impose boundary condition $\rho_{x-}=0$ since we
require that the condensate appears spontaneously.

Interestingly, we note that the equations of motion
(\ref{SLRhoxEoM}) and (\ref{SLPhiEoM}) have the useful scaling
symmetries
\begin{eqnarray}
r\rightarrow\alpha r\,,\hspace{0.5cm}(t, \varphi, x,
y)\rightarrow\frac{1}{\alpha}(t, \varphi, x,
y)\,,\hspace{0.5cm}(\rho_{x},\phi)\rightarrow\alpha(\rho_{x},\phi)\,,\hspace{0.5cm}
\label{SLsymmetry-1}
\end{eqnarray}
where $\alpha$ is a real positive number. Using these symmetries, we
can get the transformation of the relevant quantities
\begin{eqnarray}
\mu\rightarrow\alpha \mu\,,\hspace{0.5cm}\rho\rightarrow\alpha^{3}
\rho\,,\hspace{0.5cm}\rho_{x+}\rightarrow\alpha^{1+\Delta_{+}}\rho_{x+}\,.\hspace{0.5cm}
\label{SLsymmetry-2}
\end{eqnarray}
For simplicity, we will scale $r_{s}=1$ and set $q=1$ in the
following just as in \cite{Nishioka-Ryu-Takayanagi}.

\begin{figure}[ht]
\includegraphics[scale=0.65]{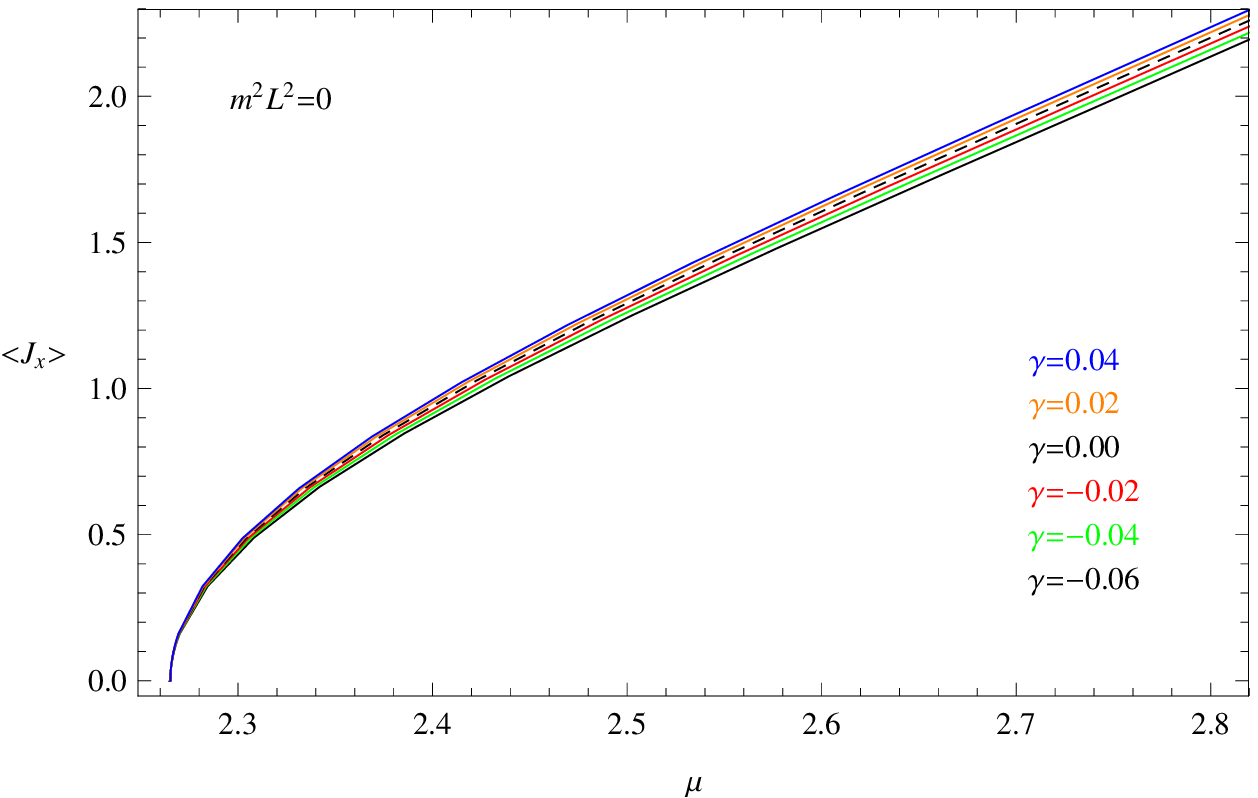}\vspace{0.0cm}
\includegraphics[scale=0.65]{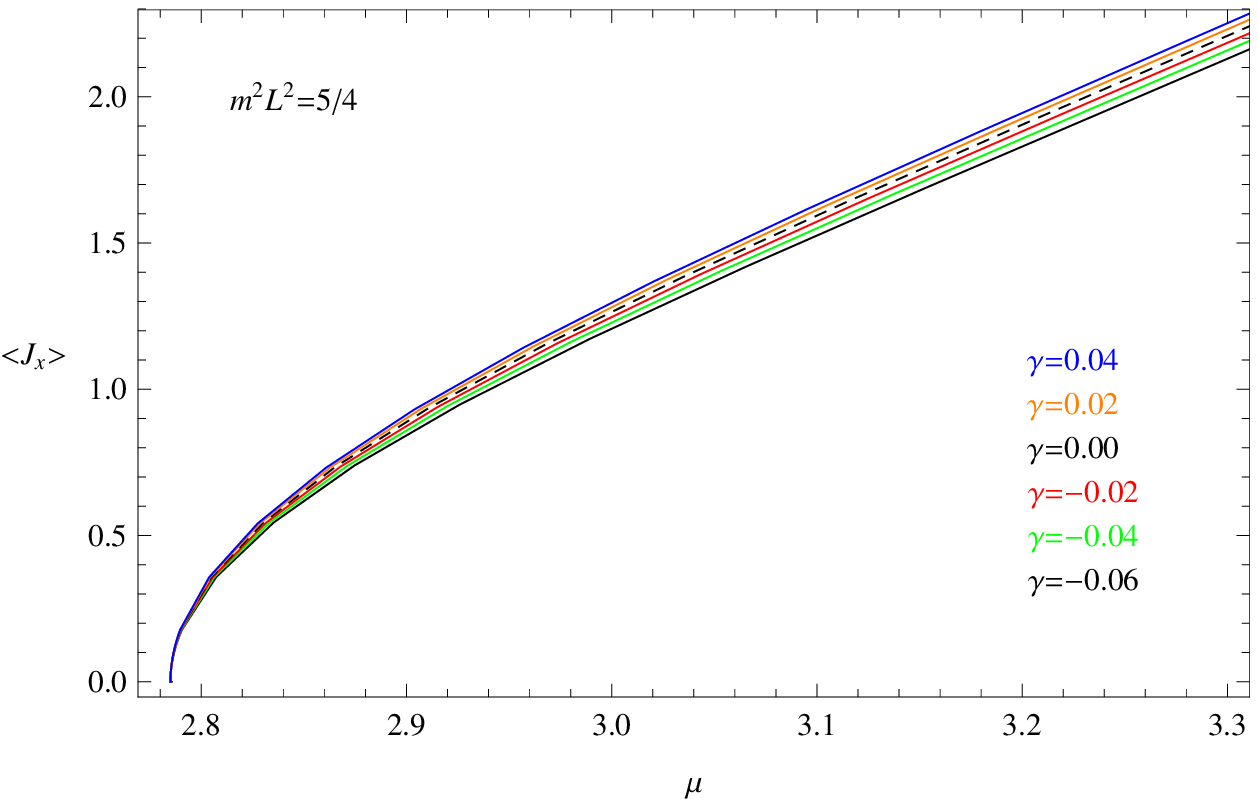}\\ \vspace{0.0cm}
\caption{\label{PWaveSLCond} (color online) The condensate of the
operator $\langle J_{x}\rangle=\rho_{x+}$ with respect to the
chemical potential $\mu$ for different Weyl coupling parameters
$\gamma$ with fixed masses of the vector field $m^{2}L^{2}=0$ (left)
and $m^{2}L^{2}=5/4$ (right) in the holographic p-wave insulator and
superconductor model. In each panel, the six lines from left to
right correspond to decreasing $\gamma$, i.e., $0.04$ (blue), $0.02$
(orange), $0$ (black and dashed), $-0.02$ (red), $-0.04$ (green) and
$\gamma=-0.06$ (black) respectively.}
\end{figure}

\begin{figure}[ht]
\includegraphics[scale=0.65]{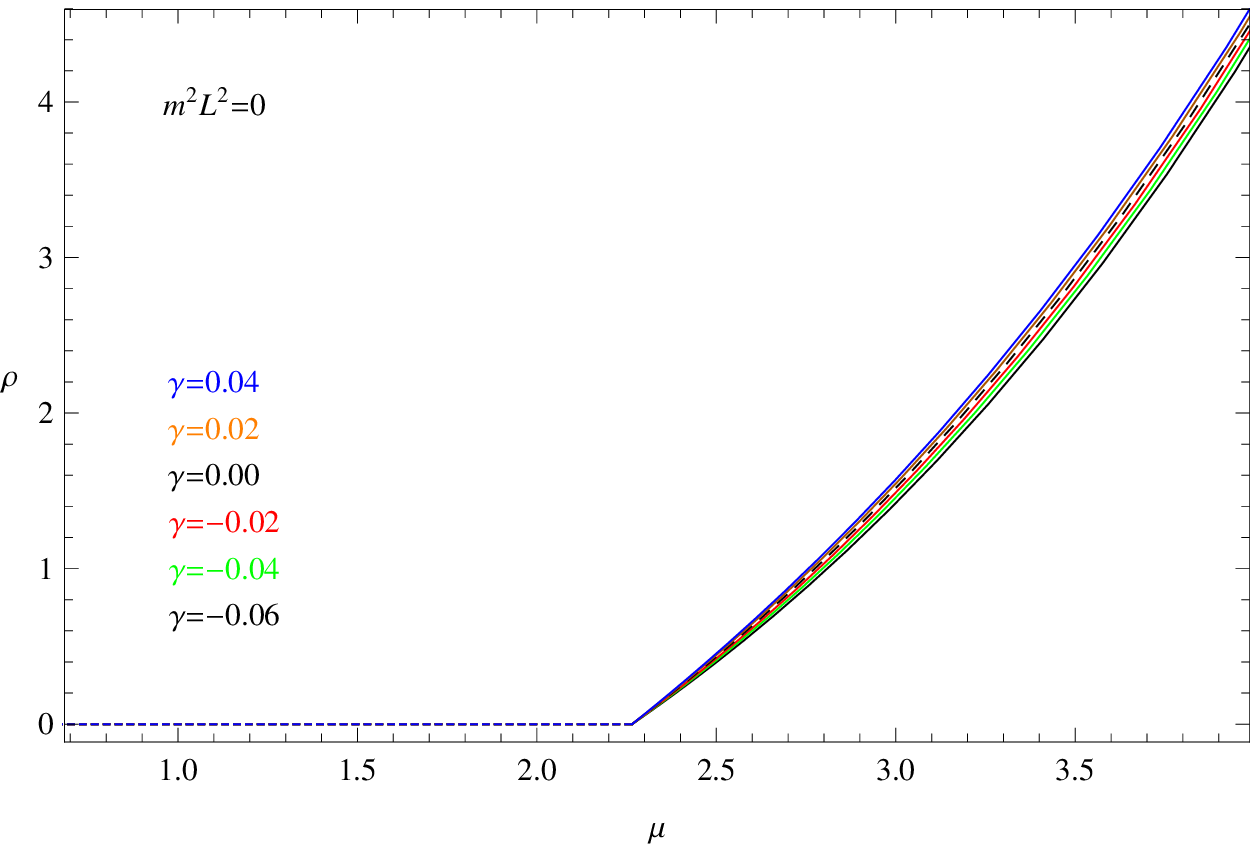}\vspace{0.0cm}
\includegraphics[scale=0.65]{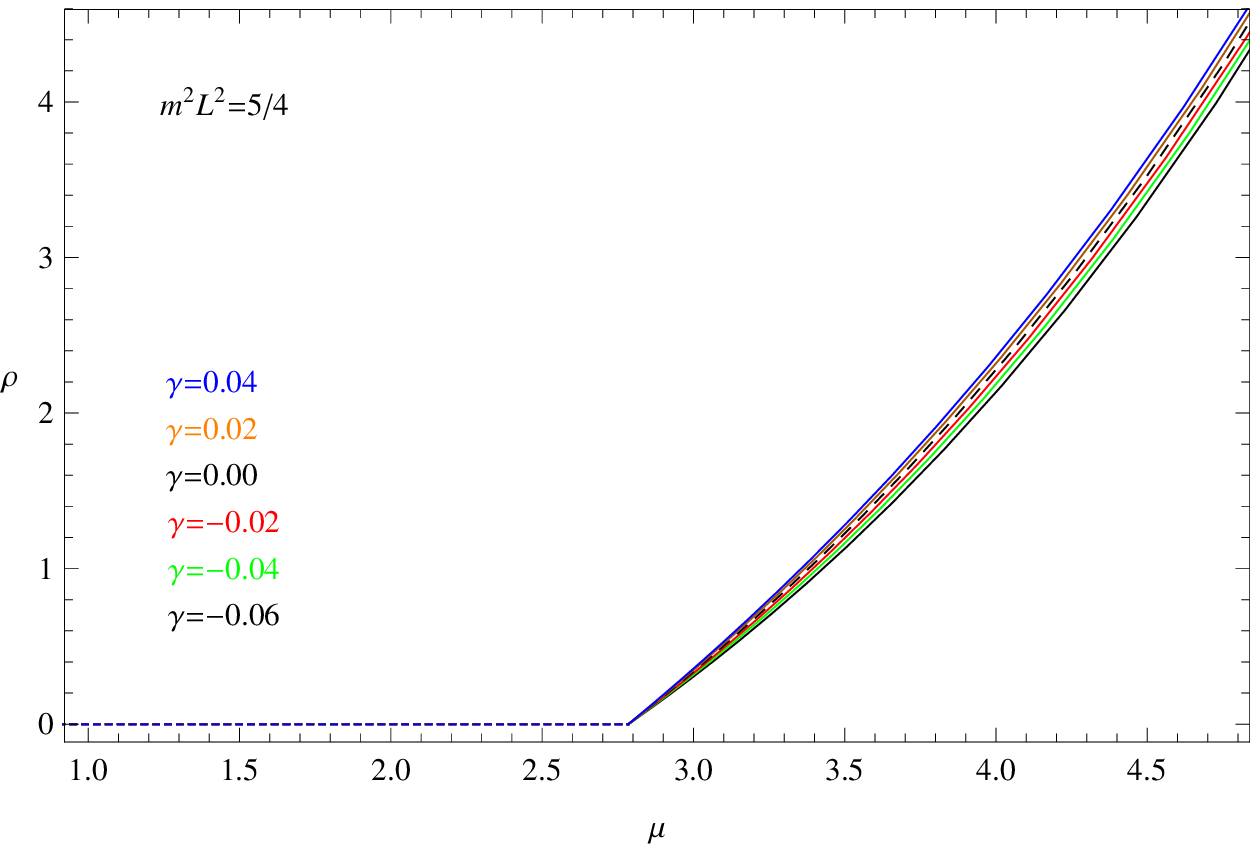}\\ \vspace{0.0cm}
\caption{\label{PWaveSLRL} (color online) The charge density $\rho$
with respect to the chemical potential $\mu$ for different Weyl
coupling parameters $\gamma$ with fixed masses of the vector field
$m^{2}L^{2}=0$ (left) and $m^{2}L^{2}=5/4$ (right) in the
holographic p-wave insulator and superconductor model. In each
panel, the six lines from left to right correspond to decreasing
$\gamma$, i.e., $0.04$ (blue), $0.02$ (orange), $0$ (black and
dashed), $-0.02$ (red), $-0.04$ (green) and $\gamma=-0.06$ (black)
respectively.}
\end{figure}

In Figs. \ref{PWaveSLCond} and \ref{PWaveSLRL} we plot the
condensate of the vector operator $\langle J_{x}\rangle$ and charge
density $\rho$ as a function of the chemical potential $\mu$ for
different Weyl coupling parameters $\gamma$ with fixed masses of the
vector field $m^{2}L^{2}=0$ (left) and $m^{2}L^{2}=5/4$ (right) in
the holographic p-wave insulator and superconductor model. From
Figs. \ref{PWaveSLCond} and \ref{PWaveSLRL}, we find that the system
is described by the AdS soliton solution itself when $\mu$ is small,
which can be interpreted as the insulator phase
\cite{Nishioka-Ryu-Takayanagi}. However, there is a second order
phase transition when $\mu\rightarrow\mu_{c}$ and the AdS soliton
reaches the superconductor (or superfluid) phase for larger $\mu$.
For the fixed Weyl coupling parameter $\gamma$, with the increase of
the vector field mass, the critical chemical potential $\mu_{c}$
becomes larger. This property agrees well with the findings in the
s-wave holographic insulator and superconductor model
\cite{ZPJ2012,Pan-Wang}. But for the fixed mass of the vector field,
it is interesting to note that the critical chemical potential
$\mu_{c}$ is independent of the Weyl coupling parameter $\gamma$,
i.e.,
\begin{eqnarray}
&&\mu_{c}=2.265,\quad {\rm for}~~m^{2}L^{2}=0~~{\rm
and}~~\forall\gamma,  \nonumber \\
&&\mu_{c}=2.785,\quad {\rm for}~~m^{2}L^{2}=5/4~~{\rm
and}~~\forall\gamma, \label{SolitonCCP}
\end{eqnarray}
which shows that the Weyl couplings will not affect the properties
of the holographic insulator/superconductor phase transition for the
fixed mass of the vector field. This behavior is reminiscent of that
seen for the holographic s-wave insulator/superconductor phase
transition with Weyl corrections, but different from the holographic
p-wave case with Weyl corrections via the Yang-Mills theory where
the corrections do have effects on the insulator/superconductor
phase transition \cite{ZPJ2012}. Thus, we conclude that the Weyl
corrections have completely different effects on the critical
chemical potential for the p-wave phase transitions of the Maxwell
complex vector field model and that of the Yang-Mills theory.

\subsection{Analytical understanding of holographic insulator/superconductor phase transition}

Since the analytic Sturm-Liouville (S-L) method, which was first
proposed by Siopsis and Therrien \cite{Siopsis} and later
generalized to study holographic insulator/superconductor phase
transition in \cite{Cai-Li-Zhang}, can clearly present the
condensation and critical phenomena of the system at the critical
point, we will apply it to investigate analytically the properties
of holographic p-wave insulator/superconductor phase transition with
Weyl corrections. In addition to back up numerical results, we will
calculate analytically the critical exponent of the system at the
critical point and obtain an analytical understanding in parallel.

Introducing the variable $z=r_{s}/r$, we can rewrite the equations
of motion (\ref{SLRhoxEoM}) and (\ref{SLPhiEoM}) into
\begin{eqnarray}
\rho_{x}^{\prime\prime}+\left(\frac{1}{z}+
\frac{f^\prime}{f}\right)\rho_{x}^\prime
+\left(\frac{\phi^2}{z^2f}-\frac{m^2}{z^4f}\right)\rho_x=0\,,
\label{SLRhoxEoMz}
\end{eqnarray}
\begin{eqnarray}
(1+8\gamma z^{4})\phi^{\prime\prime}+\left[(1+40\gamma
z^4)\frac{1}{z}+(1+8\gamma
z^{4})\frac{f^\prime}{f}\right]\phi^\prime-\frac{2\rho_x^2}{z^2f}\phi=0,
\label{SLPhiEoMz}
\end{eqnarray}
where the function $f$ now is $f(z)=(1-z^{4})/z^2$ and the prime
denotes the derivative with respect to $z$.

At the critical chemical potential $\mu_{c}$, the vector field
$\rho_x=0$. So below the critical point Eq. (\ref{SLPhiEoMz})
reduces to
\begin{eqnarray}
(1+8\gamma z^{4})\phi^{\prime\prime}+\left[(1+40\gamma
z^4)\frac{1}{z}+(1+8\gamma
z^{4})\frac{f^\prime}{f}\right]\phi^\prime=0,
\label{SLPhiEoMzCritical}
\end{eqnarray}
which results in a general solution
\begin{eqnarray}
\phi=\mu+\frac{c_{1}}{4+32\gamma}\left[4\sqrt{2\gamma}{\rm
ArcTan}(2\sqrt{2\gamma}z^{2})+\ln(1+z^2)-\ln(1-z^{2})\right],
\label{SLPhiEoMzCSolution}
\end{eqnarray}
where $c_{1}$ is an integration constant. We see that the term in
the square bracket is divergent at the tip $z=1$. Considering the
Neumann-like boundary condition (\ref{SLBoundary}) for the gauge
field $\phi$ at the tip $z=1$, we will set $c_{1}=0$ to keep $\phi$
finite, i.e., in this case $\phi$ has to be a constant. Thus, we can
get the physical solution $\phi(z)=\mu$ to Eq.
(\ref{SLPhiEoMzCritical}) if $\mu<\mu_{c}$, which agrees with our
previous numerical results.

As $\mu\rightarrow\mu_{c}$ from below the critical point, the vector
field equation (\ref{SLRhoxEoMz}) becomes
\begin{eqnarray}
\rho_{x}^{\prime\prime}+\left(\frac{1}{z}+
\frac{f^\prime}{f}\right)\rho_{x}^\prime
+\left(\frac{\mu^2}{z^2f}-\frac{m^2}{z^4f}\right)\rho_x=0\,.
\label{SRhozCriticalEoM}
\end{eqnarray}
Obviously, the Weyl coupling parameter $\gamma$ is absent in the
master Eq. (\ref{SRhozCriticalEoM}) although Eq.
(\ref{SLPhiEoMzCritical}) for the gauge field $\phi$ depends on
$\gamma$, which leads that the Weyl corrections do not have any
effect on the critical chemical potential $\mu_{c}$ for the fixed
mass of the vector field, just as shown in Figs. \ref{PWaveSLCond}
and \ref{PWaveSLRL}.

Defining a trial function $F(z)$ near the boundary $z=0$ as
\cite{Siopsis}
\begin{eqnarray}\label{SLFz}
\rho_x(z)\sim \langle J_{x}\rangle z^{\Delta_+}F(z),
\end{eqnarray}
with the boundary conditions $F(0)=1$ and $F'(0)=0$, from Eq.
(\ref{SRhozCriticalEoM}) we can get the equation of motion for
$F(z)$
\begin{eqnarray}\label{SLFzmotion}
(MF^{\prime})^{\prime}+M\left(U+\mu^2V\right)F=0,
\end{eqnarray}
where we have introduced
\begin{eqnarray}\label{MUVFu}
M=z^{2\Delta_{+}-1}(z^{4}-1),~~
U=\frac{\Delta_{+}(\Delta_{+}-1)}{z^{2}}+\frac{\Delta_{+}}{z}\left(\frac{1}{z}+\frac{f'}{f}\right)-\frac{m^2}{z^{4}f},~~
V=\frac{1}{z^{2}f}.
\end{eqnarray}
Following the S-L eigenvalue problem \cite{Gelfand-Fomin}, we obtain
the expression which can be used to estimate the minimum eigenvalue
of $\mu^2$
\begin{eqnarray}\label{SLEigenvalue}
\mu^2=\frac{\int^{1}_{0}M\left(F'^{2}-UF^{2}\right)dz}{\int^{1}_{0}MVF^{2}dz}=\frac{\Xi(a,m)}{\Sigma(a,m)},
\end{eqnarray}
with
\begin{eqnarray}
\Xi(a,m)&=&\left(a-1\right)^2+\frac{1+m^2}{2\left(\Delta_+-1\right)}-\frac{am^2}{\Delta_+}+
\frac{a^2\left(1+m^2\right)-1}{2\left(1+\Delta_+\right)}-\frac{a^2}{2\left(3+\Delta_+\right)},
\nonumber\\
\Sigma(a,m)&=&\frac{1}{2}\left[\frac{1}{\Delta_+}+a\left(\frac{a}{2+\Delta_+}-\frac{2}{1+\Delta_+}\right)\right],
\end{eqnarray}
where we have assumed the trial function to be $F(z)=1-az^{2}$ with
a constant $a$ in the calculation. For different values of the mass
of the vector field, we can get the minimum eigenvalue of $\mu^{2}$
and the corresponding value of $a$, for example,
$\mu_{min}^{2}=7.768$ and $a=0.382$ for $m^{2}L^{2}=5/4$, which lead
to the critical chemical potential $\mu_{c}=\mu_{min}=2.787$. In
Table \ref{SLTable}, we present the critical chemical potential
$\mu_{c}$ for chosen mass of the vector field. Comparing with
numerical results, we observe that the analytic results derived from
S-L method are in very good agreement with the numerical
computations.

\begin{table}[ht]
\begin{center}
\caption{\label{SLTable} The critical chemical potential $\mu_{c}$
for the vector operator $\langle{J_{x}}\rangle$ obtained by the
analytical S-L method and numerical shooting method with chosen
various masses of the vector field for the holographic p-wave
insulator/superconductor model. It should be noted that the Weyl
corrections do not have any effect on $\mu_{c}$ for the fixed mass
of the vector field.}
\begin{tabular}{c c c c c c c c c}
         \hline \hline
$m^{2}L^{2}$ & -1/2 & -1/4 & 0 & 1/4 & 1/2 & 3/4 & 1 & 5/4
        \\
        \hline
~~~~Analytical~~~~&~~~~$1.959$~~~~&~~~~$2.127$~~~~&~~~~$2.267$~~~~
&~~~~$2.390$~~~~&~~~~$2.502$~~~~&~~~~$2.603$~~~~&~~~~$2.698$~~~~&~~~~$2.787$~~~~
          \\
~~~~Numerical~~~~&~~~~$1.958$~~~~&~~~~$2.125$~~~~&~~~~$2.265$~~~~
&~~~~$2.388$~~~~&~~~~$2.500$~~~~&~~~~$2.601$~~~~&~~~~$2.696$~~~~&~~~~$2.785$~~~~
          \\
        \hline \hline
\end{tabular}
\end{center}
\end{table}

From Table \ref{SLTable}, we find that, with the increase of the
mass of the vector field, the critical chemical potential $\mu_{c}$
becomes larger, which agrees with our previous numerical results.
More importantly, due to the absence of the Weyl coupling parameters
from the master Eq. (\ref{SRhozCriticalEoM}), the Weyl corrections
do not have any effect on the critical chemical potential $\mu_{c}$
for the fixed mass of the vector field, which supports the numerical
finding as shown in Figs. \ref{PWaveSLCond} and \ref{PWaveSLRL}.

Now we are in a position to study the critical phenomena of this
holographic p-wave system. Noting that the condensation of the
vector operator $\langle J_{x}\rangle$ is so small when
$\mu\rightarrow\mu_{c}$, we will expand $\phi(z)$ in small $\langle
J_{x}\rangle$ as
\begin{eqnarray}\label{SLEigenvalue}
\phi(z)\sim\mu_{c}+\langle J_{x}\rangle\chi(z)+\cdots,
\end{eqnarray}
where the boundary condition is $\chi(1)=0$ at the tip. Defining a
function $\xi(z)$ as
\begin{eqnarray}\label{SLXiz}
\chi(z)=2\mu_{c}\langle J_{x}\rangle\xi(z),
\end{eqnarray}
we will have the equation of motion for $\xi(z)$
\begin{eqnarray}\label{SLXiEoM}
(K\xi')'-\frac{z^{2\Delta_{+}-3}(z^4-1)F^2}{f}=0,
\end{eqnarray}
with
\begin{eqnarray}\label{SLQz}
K(z)=\frac{\left(z^4-1\right)\left(1+8\gamma z^4\right)}{z}.
\end{eqnarray}

According to the asymptotic behavior in Eq. (\ref{SLinfinity}), we
can expand $\phi$ near $z\rightarrow0$ as
\begin{eqnarray}\label{SLPhiExpand}
\phi(z)\simeq\mu-\rho z^2\simeq\mu_c +\langle
J_{x}\rangle\left[\chi(0)+\chi^\prime(0)z+\frac{1}{2}\chi^{\prime\prime}(0)z^2+\cdot\cdot\cdot\right].
\end{eqnarray}
From the coefficients of the $z^0$ term in both sides of the above
formula and with the help of Eq. (\ref{SLXiz}), we arrive at
\begin{eqnarray}\label{SLJxExp}
\langle
J_{x}\rangle=\frac{1}{\left[2\mu_c\xi(0)\right]^{\frac{1}{2}}}\left(\mu-\mu_c\right)^{\frac{1}{2}},
\end{eqnarray}
with
\begin{eqnarray}
\xi(0)=c_2-\int^1_0\frac{1}{K(z)}\left[c_3+\int^z_1\frac{
x^{2\Delta_{+}-3}F(x)^2(x^{4}-1)}{f(x)}dx\right]dz,
\end{eqnarray}
where the integration constants $c_2$ and $c_3$ can be determined by
the boundary condition $\chi(z)$. For example, for the case of
$\gamma=0.02$ with $m^{2}L^{2}=5/4$, we have $\langle
J_{x}\rangle\approx1.891(\mu-\mu_{c})^{1/2}$ when $a=0.382$, which
agrees well with the numerical result given in the right panel of
Fig. \ref{PWaveSLCond}. Note that the expression (\ref{SLJxExp}) is
valid for all cases considered here. Thus, the vector operator
$\langle J_{x}\rangle$ satisfies $\langle
J_{x}\rangle\sim(\mu-\mu_{c})^{1/2}$ near the critical point, which
holds for various values of Weyl coupling parameters and masses of
the vector field. The analytic result shows that the holographic
p-wave insulator/superconductor phase transition belongs to the
second order and the critical exponent of the system takes the
mean-field value $1/2$, which can be used to back up the numerical
findings obtained from Fig. \ref{PWaveSLCond}.

Comparing the coefficients of the $z^1$ term in Eq.
(\ref{SLPhiExpand}), we see that $\chi^\prime(0)\rightarrow 0$,
which leads to $\xi^\prime(0)\rightarrow 0$. This behavior is
consistent with the following relation by making integration of both
sides of Eq. (\ref{SLXiEoM})
\begin{eqnarray}
\left[\frac{\xi'(z)}{z}\right]\bigg|_{z\rightarrow
0}=\int_{0}^{1}\frac{z^{2\Delta_{+}-3}(z^4-1)F^2}{f}dz.
\end{eqnarray}

Considering the coefficients of the $z^2$ term in Eq.
(\ref{SLPhiExpand}), we get
\begin{eqnarray}\label{SLRhoExp}
\rho=-\frac{1}{2}\langle
J_{x}\rangle\chi''(0)=\Gamma(\gamma,m)(\mu-\mu_{c}),
\end{eqnarray}
with
\begin{eqnarray}
\Gamma(\gamma,m)=-\frac{1}{2\xi(0)}\int_{0}^{1}\frac{z^{2\Delta_{+}-3}(z^4-1)F^2}{f}dz,
\end{eqnarray}
which is a function of the Weyl coupling parameter and the vector
field mass. For the case of $\gamma=0.02$ with $m^{2}L^{2}=5/4$, as
an example, we can find $\rho=1.068\left(\mu-\mu_c\right)$ when
$a=0.382$, which is in good agreement with the result shown in the
right panel of Fig. \ref{PWaveSLRL}. Since the Weyl coupling
parameters and masses of the vector field will not alter Eq.
(\ref{SLRhoExp}), we can obtain the linear relation between the
charge density and the chemical potential near $\mu_{c}$, i.e.,
$\rho\sim(\mu-\mu_{c})$, which supports the numerical result
presented in Fig. \ref{PWaveSLRL}.

\section{p-wave superconductor models with Weyl corrections in AdS black hole}

Since the Weyl couplings will not affect the properties of the new
p-wave superconductor in AdS soliton via the Maxwell complex vector
field model, which is different from that via the Yang-Mills theory
where the Weyl corrections do have effects on the
insulator/superconductor phase transition, it seems to be an
interesting study to consider the influences of the Weyl corrections
on this new p-wave superconductor in AdS black hole.

\subsection{Numerical investigation of holographic metal/superconductor phase transition}

In the probe limit, the background metric is a five-dimensional
planar Schwarzschild-AdS black hole
\begin{eqnarray}\label{SchBlackhole}
ds^2=-f\left(r\right)dt^2+\frac{1}{f\left(r\right)}dr^2+r^2\left(dx^2+dy^2+dz^2\right),
\end{eqnarray}
where $f(r)=r^2(1-r_{+}^{4}/r^{4})$ with the radius of the event
horizon $r_{+}$. The Hawking temperature of the black hole can be
expressed as
\begin{eqnarray}
T=\frac{r_{+}}{\pi},
\end{eqnarray}
which can be interpreted as the temperature of the CFT. The metric
(\ref{SchBlackhole}) has the following nonzero components of the
Weyl tensor $C_{\mu\nu\rho\sigma}$
\begin{eqnarray}
C_{0i0j}=\frac{r_+^4f\left(r\right)}{r^2}\delta_{ij}, \quad
C_{0r0r}=-\frac{3r_+^4}{r^4} ,\quad
C_{irjr}=-\frac{r_+^4}{r^2f\left(r\right)}\delta_{ij},\quad
C_{ijkl}=r_+^4\delta_{ik}\delta_{jl},
\end{eqnarray}
with $i,~j,~k,~l=x$, $y$ or $z$.

For completeness, we still work on the ansatz (\ref{Ansatz}) and get
the equations of motion from the action (\ref{NewPWction}) in the
Maxwell complex vector field model
\begin{eqnarray}\label{BRhorEoM}
\rho_{x}^{\prime\prime}+\left(
\frac{1}{r}+\frac{f^\prime}{f}\right)\rho_{x}^\prime
+\left(\frac{q^2 \phi^2}{f^2}-\frac{m^2}{f}\right)\rho_{x}=0\,,
\end{eqnarray}
\begin{eqnarray}\label{BPhirEoM}
\left(1-\frac{24\gamma
r_+^4}{r^4}\right)\phi^{\prime\prime}+\frac{3}{r}\left(1+\frac{8\gamma
r_+^4}{r^4}\right)\phi^\prime-\frac{2q^2\rho_x^2}{r^2f}\phi=0,
\end{eqnarray}
where the prime denotes the derivative with respect to $r$.

In order to solve the equations of motion (\ref{BRhorEoM}) and
(\ref{BPhirEoM}) numerically, we have to impose the appropriate
boundary conditions for $\rho_{x}(r)$ and $\phi(r)$. At the horizon
$r=r_{+}$, the boundary conditions are
\begin{eqnarray}
\rho_{x}(r_{+})=\frac{f^\prime(r_{+})}{m^{2}}\rho_{x}^\prime(r_{+})\,,\hspace{0.5cm}
\phi(r_{+})=0\,. \label{BHhorizon}
\end{eqnarray}
Obviously, we require $\phi\left(r_+\right)=0$ in order for
$g^{\mu\nu}A_\mu A_\nu$ to be finite at the horizon, which is in
strong contrast to that of the AdS soliton where there is a constant
nonzero gauge field $\phi(r_{s})$ at $r=r_{s}$. But near the
boundary $r\rightarrow\infty$, we find that the solutions have the
same boundary conditions just as Eq. (\ref{SLinfinity}) for the
holographic p-wave insulator and superconductor model with Weyl
corrections.

For the equations of motion (\ref{BRhorEoM}) and (\ref{BPhirEoM}),
we can also obtain the useful scaling symmetries
\begin{eqnarray}
r\rightarrow\alpha
r\,,\hspace{0.5cm}(t,x,y,z)\rightarrow\frac{1}{\alpha}(t,x,y,z)\,,\hspace{0.5cm}(\rho_{x},\phi)\rightarrow\alpha(\rho_{x},\phi)\,,\hspace{0.5cm}
\label{BHsymmetry-1}
\end{eqnarray}
which result in the transformation of the relevant quantities
\begin{eqnarray}
T\rightarrow\alpha T\,,\hspace{0.5cm}\mu\rightarrow\alpha
\mu\,,\hspace{0.5cm}\rho\rightarrow\alpha^{3}
\rho\,,\hspace{0.5cm}\rho_{x+}\rightarrow\alpha^{1+\Delta_{+}}\rho_{x+}\,,\hspace{0.5cm}
\label{BHsymmetry-2}
\end{eqnarray}
with a real positive number $\alpha$. Without loss of generality, we
can scale $r_{+}=1$ and set $q=1$ in the following just as in
\cite{CaiPWave-1}.

\begin{figure}[ht]
\includegraphics[scale=0.65]{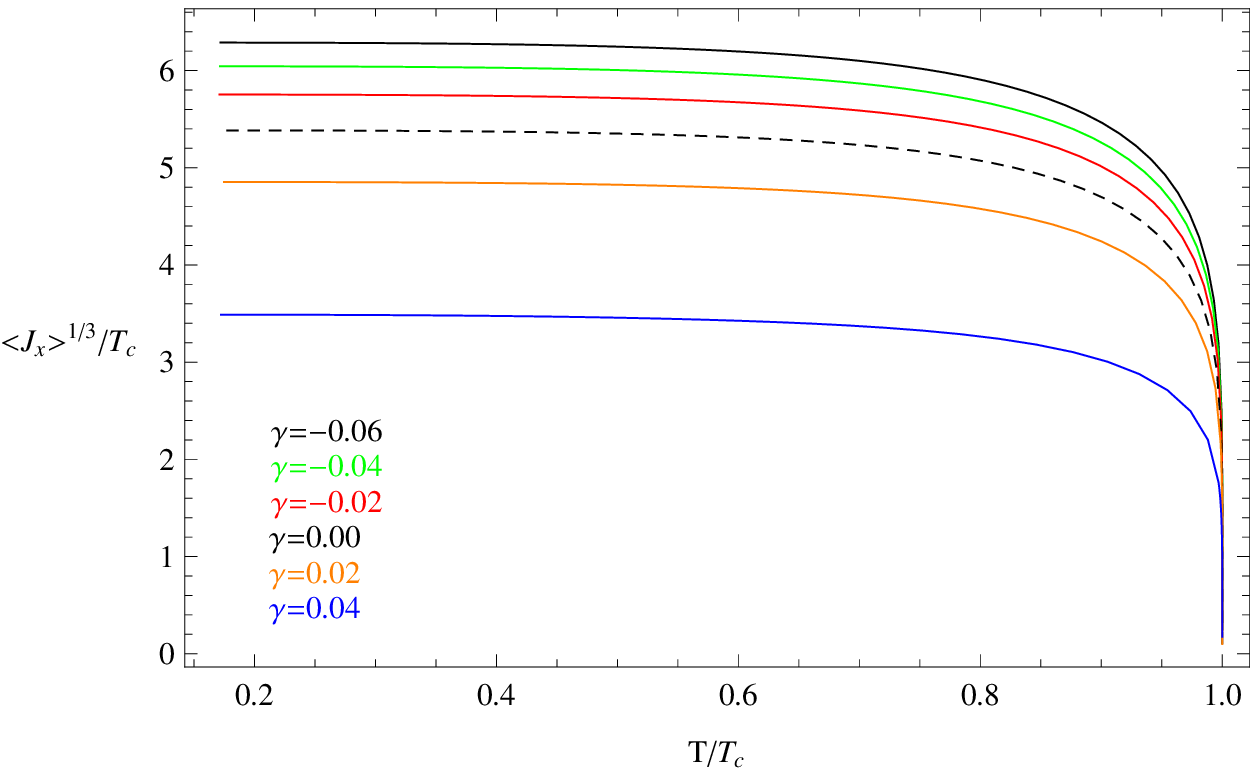}\vspace{0.0cm}
\includegraphics[scale=0.65]{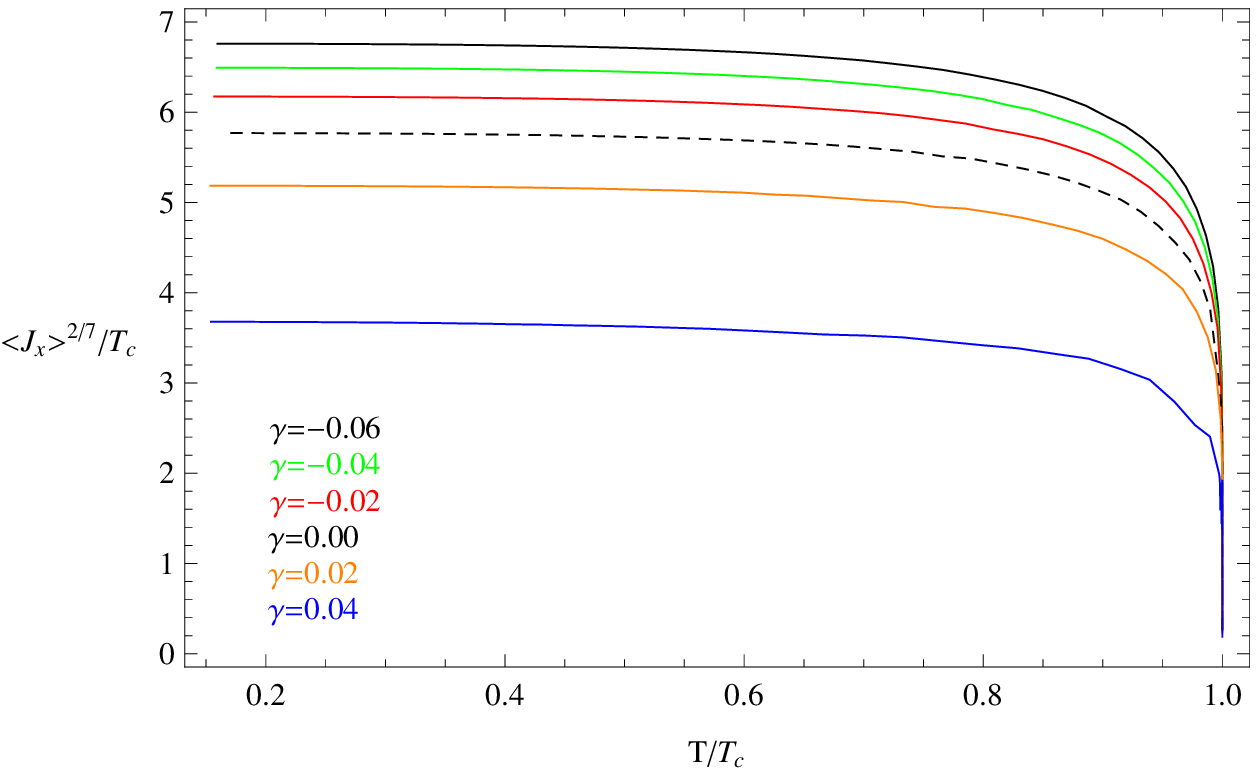}\\ \vspace{0.0cm}
\caption{\label{PWaveCondBH} (color online) The condensate of the
operator $\langle J_{x}\rangle=\rho_{x+}$ as a function of
temperature for different Weyl coupling parameters $\gamma$ with
fixed masses of the vector field $m^{2}L^{2}=0$ (left) and
$m^{2}L^{2}=5/4$ (right) in the holographic p-wave superconductor
model. In each panel, the six lines from top to bottom correspond to
increasing $\gamma$, i.e., $\gamma=-0.06$ (black), $-0.04$ (green),
$-0.02$ (red), $0$ (black and dashed), $0.02$ (orange) and $0.04$
(blue) respectively.}
\end{figure}

In Fig. \ref{PWaveCondBH}, we present the condensate of the vector
operator $\langle J_{x}\rangle$ as a function of temperature for
different Weyl coupling parameters $\gamma$ with fixed masses of the
vector field $m^{2}L^{2}=0$ (left) and $m^{2}L^{2}=5/4$ (right) in
the holographic p-wave superconductor model. Obviously, we observe
that the behavior of each curve for the fixed $\gamma$ and
$m^{2}L^{2}$ is in good agreement with the holographic
superconducting phase transition in the literature, which shows that
the black hole solution with non-trivial vector field can describe a
superconducting phase.

From Fig. \ref{PWaveCondBH}, for the fixed Weyl coupling parameter
$\gamma$, we find that the condensation gap for the vector operator
$\langle J_{x}\rangle$ becomes larger with the increase of the mass
of the vector field, which implies that the increase of the mass
makes it harder for the vector operator to condense. However, if we
concentrate on the same mass of the vector field, we see that the
higher correction term $\gamma$ makes the condensation gap smaller,
which means that the condensation is easier to be formed when the
parameter $\gamma$ increases. In fact, the table \ref{Tc-D5} shows
that the critical temperature $T_{c}$ for the vector operator
$\langle J_{x}\rangle$ with the fixed vector field mass increases as
the correction term $\gamma$ increases, which agrees well with the
finding in Fig. \ref{PWaveCondBH}. This behavior is reminiscent of
that seen for the holographic p-wave dual models via the Yang-Mills
theory, where the critical temperature $T_{c}$ increases as the Weyl
correction $\gamma$ increases \cite{MomeniSL}. So we conclude that
these two p-wave models with Weyl corrections share some similar
features for the condensation of the vector operator.

\begin{table}[ht]
\caption{\label{Tc-D5} The critical temperature $T_{c}$ obtained by
the numerical shooting method for the vector operator $\langle
J_{x}\rangle$ with different Weyl coupling parameters $\gamma$ and
fixed masses of the vector field, i.e., $m^2L^2=0$ and $5/4$. We
have set $\rho=1$ in the table.}
\begin{tabular}{c c c c c c c}
         \hline
$\gamma$ & -0.06 & -0.04 & -0.02 & 0 & 0.02 & 0.04
        \\
        \hline
~~~$m^2L^2=0$~~~~~~&~~~~~~$0.172$~~~~~~&~~~~~~$0.179$~~~~~~&~~~~~~$0.188$~~~~~~&~~~~~~$0.201$~~~~~~&~~~~~~$0.222$~~~~~~&~~~~~~$0.309$~~~
          \\
~~~$m^2L^2=5/4$~~~~~~&~~~~~~$0.157$~~~~~~&~~~~~~$0.163$~~~~~~&~~~~~~$0.172$~~~~~~&~~~~~~$0.184$~~~~~~&~~~~~~$0.205$~~~~~~&~~~~~~$0.288$~~~
          \\
        \hline
\end{tabular}
\end{table}

\subsection{Analytical understanding of holographic metal/superconductor phase transition}

We still use the S-L method to deal with the effect of the Weyl
corrections on the holographic p-wave metal/superconductor phase via
the Maxwell complex vector field model. Changing the coordinate and
setting $z=r_{+}/r$, we can convert the equations of motion
(\ref{BRhorEoM}) and (\ref{BPhirEoM}) to be
\begin{eqnarray}
\rho_{x}^{\prime\prime}+\left(\frac{1}{z}+
\frac{f^\prime}{f}\right)\rho_{x}^\prime
+\frac{1}{z^4}\left(\frac{\phi^2}{r_{+}^2f^2}-\frac{m^2}{f}\right)\rho_x=0\,,
\label{BRhozEoM}
\end{eqnarray}
\begin{eqnarray}
\left(1-24\gamma
z^{4}\right)\phi^{\prime\prime}-\left(\frac{1}{z}+72\gamma
z^3\right)\phi^\prime-\frac{2\rho_x^2}{r_{+}^2z^2f}\phi=0.
\label{BPhizEoM}
\end{eqnarray}
Here the function $f$ has been rewritten into $f(z)=(1-z^{4})/z^2$
and the prime denotes the derivative with respect to $z$.

At the critical temperature $T_{c}$, the vector field $\rho_x=0$.
Thus, below the critical point Eq. (\ref{BPhizEoM}) becomes
\begin{eqnarray}
\left(1-24\gamma
z^{4}\right)\phi^{\prime\prime}-\left(\frac{1}{z}+72\gamma
z^3\right)\phi^\prime=0. \label{BPhiCriticalEoM}
\end{eqnarray}
Note that at $z=0$, from the boundary condition we have
\begin{eqnarray}
\phi^{\prime\prime}|_{z=0}=-\frac{2\rho}{r_{+}^{2}} \quad {\rm
at}~~T=T_{c}.
\end{eqnarray}
Thus, neglecting terms of order $0(\gamma^{n\geq2})$, we can obtain
the solution to Eq. (\ref{BPhiCriticalEoM})
\begin{eqnarray}
\phi(z)=\lambda r_{+}\varphi_{1}(z)=\lambda
r_{+}\left(1-z^2\right)\left[1+8\gamma\left(1+z^2+z^4\right)\right],
\label{BPhiCriticalSolution}
\end{eqnarray}
with $\lambda=\rho/r_{+}^3$.

Near the boundary $z=0$, we introduce a trial function $F(z)$
\begin{eqnarray}
\rho_x|\sim\frac{\langle J_{x}\rangle}{r_{+}^{\Delta_{+}}}
z^{\Delta_{+}}F(z), \label{BintroduceF}
\end{eqnarray}
with the boundary conditions $F(0)=1$ and $F'(0)=0$. Therefore the
equation of motion for $F(z)$ is given by
\begin{eqnarray}\label{BFEoM}
(MF^{\prime})^{\prime}+M\left(P+\lambda^2Q\right)F=0,
\end{eqnarray}
with
\begin{eqnarray}
P=\frac{\Delta_{+}}{z}\left(\frac{\Delta_{+}}{z}+\frac{f'}{f}\right)-\frac{m^2}{z^{4}f},~~
Q=\frac{\varphi_{1}^{2}}{z^{4}f^{2}},
\end{eqnarray}
where $M(z)$ has been defined in (\ref{MUVFu}). According to the S-L
eigenvalue problem \cite{Gelfand-Fomin}, we deduce the eigenvalue
$\lambda$ minimizes the expression
\begin{eqnarray}\label{lambdaeigenvalue}
\lambda^{2}=\frac{\int^{1}_{0}M\left(F'^{2}-PF^{2}\right)dz}{\int^{1}_{0}MQF^2dz}.
\end{eqnarray}
Here we still assume the trial function to be $F(z)=1-az^{2}$ with a
constant $a$. Using above equation to compute the minimum eigenvalue
of $\lambda^{2}$, we can get the critical temperature $T_c$ for
different Weyl coupling parameters $\gamma$ and masses of the vector
field $m$ from the following relation
\begin{eqnarray}\label{TcRelation}
T_c=\frac{1}{\pi\lambda_{min}^{1/3}}\sqrt[3]{\rho}.
\end{eqnarray}
As an example, for the case of $m^{2}L^{2}=5/4$ with the chosen
value of the Weyl coupling parameter $\gamma=0.02$, we obtain the
minimum $\lambda_{min}=18.069$ at $a=0.748$. According to the
relation (\ref{TcRelation}), we can easily get the critical
temperature $T_{c}=0.196\sqrt[3]{\rho}$, which is consistent with
the numerical result $T_{c}=0.205\sqrt[3]{\rho}$ in Table
\ref{Tc-D5}. In Table \ref{Tc-D5Analytical} we give the critical
temperature $T_c$ obtained by the analytical S-L method for the
vector operator $\langle J_{x}\rangle$ when we fix the mass of the
vector field $m^2L^2$ for different Weyl couplings by choosing the
expanded solution (\ref{BPhiCriticalSolution}) upto first order in
the Weyl coupling parameter. Comparing with the numerical results of
Table \ref{Tc-D5} in the range $-0.02\leq\gamma\leq0.02$, we observe
that the differences between the analytical and numerical values are
within $4.5\%$.

\begin{table}[ht]
\caption{\label{Tc-D5Analytical} The critical temperature $T_{c}$
obtained by the analytical S-L method for the vector operator
$\langle J_{x}\rangle$ with different Weyl coupling parameters
$\gamma$ and fixed masses of the vector field, i.e., $m^2L^2=0$ and
$5/4$. We have set $\rho=1$ and chosen different expanded solutions
to Eq. (\ref{BPhiCriticalEoM}), i.e., $\varphi_{1}$ upto $0(\gamma)$
and $\varphi_{2}$ upto $0(\gamma^{6})$ in the table.}
\begin{tabular}{c c c c c c}
         \hline
$\gamma$ & -0.02 & -0.01 & 0 & 0.01 & 0.02
        \\
        \hline
~~~$m^2L^2=0~~(\varphi_{1})$~~~~~~&~~~~~~$0.181$~~~~~~&~~~~~~$0.190$~~~~~~&~~~~~~$0.199$~~~~~~&~~~~~~$0.207$~~~~~~&~~~~~~$0.214$~~~
          \\
~~~$m^2L^2=0~~(\varphi_{2})$~~~~~~&~~~~~~$0.186$~~~~~~&~~~~~~$0.192$~~~~~~&~~~~~~$0.199$~~~~~~&~~~~~~$0.208$~~~~~~&~~~~~~$0.221$~~~
          \\
~~~$m^2L^2=5/4~~(\varphi_{1})$~~~~~~&~~~~~~$0.165$~~~~~~&~~~~~~$0.174$~~~~~~&~~~~~~$0.182$~~~~~~&~~~~~~$0.189$~~~~~~&~~~~~~$0.196$~~~
          \\
~~~$m^2L^2=5/4~~(\varphi_{2})$~~~~~~&~~~~~~$0.169$~~~~~~&~~~~~~$0.175$~~~~~~&~~~~~~$0.182$~~~~~~&~~~~~~$0.191$~~~~~~&~~~~~~$0.203$~~~
          \\
        \hline
\end{tabular}
\end{table}

When we perform analytic computation of the solution to Eq.
(\ref{BPhiCriticalEoM}) upto sixth order in the Weyl coupling
parameter $\gamma$, i.e., change the solution
(\ref{BPhiCriticalSolution}) into
\begin{eqnarray}
\phi(z)&=&\lambda r_{+}\varphi_{2}(z)\nonumber \\
&=&\lambda
r_{+}\left[\varphi_{1}(z)+\frac{24^{2}}{5}(1-z^{10})\gamma^{2}+
\frac{24^{3}}{7}(1-z^{14})\gamma^{3}+\frac{24^{4}}{9}(1-z^{18})\gamma^{4}
+\frac{24^{5}}{11}(1-z^{22})\gamma^{5}+\frac{24^{6}}{13}(1-z^{26})\gamma^{6}\right],
\nonumber \\
\label{BPhiCSNew}
\end{eqnarray}
the agreement of the analytic results presented in Table
\ref{Tc-D5Analytical} with the numerical calculation shown in Table
\ref{Tc-D5} is impressive. Thus, we can improve the analytic result
and get the critical temperature more consistent with the numerical
result if we expand the solution to Eq. (\ref{BPhiCriticalEoM}) upto
a sufficiently high order in the Weyl coupling parameter $\gamma$,
even we consider the case of larger $\gamma$.

From Table \ref{Tc-D5Analytical}, we point out that the critical
temperature $T_{c}$ increases as the Weyl correction $\gamma$
increases for the fixed vector field mass but decreases as the mass
$m^{2}$ increases for the fixed Weyl coupling parameter, which
supports the numerical computation shown in Fig. \ref{PWaveCondBH}
and Table \ref{Tc-D5}.

We will investigate the critical phenomena of the system. Since the
condensation for the vector operator $\langle J_{x}\rangle$ is so
small when $T \rightarrow T_c$, we can expand $\phi(z)$ in $\langle
J_{x}\rangle$ near $z=0$
\begin{eqnarray}\label{BPhiExpandNearTc}
\frac{\phi(z)}{r_+}=\lambda\varphi_{1}+\frac{\langle
J_{x}\rangle^2}{r_+^{2(1+\Delta_{+})}}\chi\left(z\right)+\cdot\cdot\cdot,
\end{eqnarray}
with the boundary conditions $\chi(1)=0$ and $\chi'(1)=0$
\cite{Siopsis,Li-Cai-Zhang}. Thus, substituting the functions
(\ref{BintroduceF}) and (\ref{BPhiExpandNearTc}) into
(\ref{BPhizEoM}), we can get the equation of motion for
$\chi\left(z\right)$
\begin{eqnarray}\label{BHChizEoM}
\left(W\chi^\prime\right)^\prime-\frac{2\lambda
z^{2\Delta_{+}-3}F^2\varphi_{1}}{f}=0,
\end{eqnarray}
where we have introduced a new function
\begin{eqnarray}
W(z)=\frac{1-24\gamma z^4}{z}.
\end{eqnarray}

From the asymptotic behavior (\ref{SLinfinity}), near
$z\rightarrow0$ we can arrive at
\begin{eqnarray}\label{BHPhiExpand}
\frac{\phi(z)}{r_+}&=&\frac{\rho}{r^{3}_+}(1-z^{2})\nonumber \\
&=&
\lambda\left(1-z^2\right)\left[1+8\gamma\left(1+z^2+z^4\right)\right]+\frac{\langle
J_{x}\rangle^2}{r_+^{2(1+\Delta_{+})}}\left[\chi(0)+\chi^\prime(0)z+\frac{1}{2}\chi^{\prime\prime}(0)z^2+\cdot\cdot\cdot\right].
\end{eqnarray}
Considering the coefficients of the $z^1$ term in both sides of the
above formula, we can find that $\chi^\prime(0)\rightarrow 0$, which
agrees well with the following relation by making integration of
both sides of Eq. (\ref{BHChizEoM})
\begin{eqnarray}
\left[\frac{\chi'(z)}{z}\right]\bigg|_{z\rightarrow
0}=-2\lambda\Upsilon(\gamma,m)=-2\lambda\int_{0}^{1}\frac{z^{2\Delta_{+}-3}F^2\varphi_{1}}{f}dz,
\end{eqnarray}
where $\Upsilon(\gamma,m)$ is a function of the Weyl coupling
parameter and the vector field mass. Comparing the coefficients of
the $z^2$ term in Eq. (\ref{BHPhiExpand}), we have
\begin{eqnarray}\label{BHRhoExp}
\frac{\rho}{r^{3}_{+}}=\lambda-\frac{\langle
J_{x}\rangle^2}{2r_+^{2(1+\Delta_{+})}}\chi''(0),
\end{eqnarray}
which leads to
\begin{eqnarray}\label{BHJxExp}
\langle J_{x}\rangle=\left(\pi
T_{c}\right)^{1+\Delta_{+}}\sqrt{\frac{3}{\Upsilon(\gamma,m)}}\left(1-\frac{T}{T_c}\right)^{\frac{1}{2}}.
\end{eqnarray}
Obviously, the expression (\ref{BHJxExp}) is valid for all cases
considered here. For example, for the case of $\gamma=0.02$ with
$m^{2}L^{2}=5/4$, we have $\langle
J_{x}\rangle\approx1.580(1-T/T_{c})^{1/2}$ when $a=0.748$, which is
in agreement with the numerical calculation given in the right panel
of Fig. \ref{PWaveCondBH}. Since the Weyl coupling parameters and
masses of the vector field will not alter Eq. (\ref{BHJxExp}) except
for the prefactor, we can obtain the relation $\langle
J_{x}\rangle\sim(1-T/T_{c})^{1/2}$ near the critical point. The
analytic result shows that the holographic p-wave
metal/superconductor phase transition belongs to the second order
and the critical exponent of the system takes the mean-field value
$1/2$, which can be used to back up the numerical findings shown in
Fig. \ref{PWaveCondBH}.

\section{Conclusions}

In the probe limit, we have investigated the holographic p-wave dual
models with Weyl corrections both in the backgrounds of AdS soliton
and AdS black hole in order to understand the influences of the
$1/N$ or $1/\lambda$ corrections on the vector condensate via a
Maxwell complex vector field model. Different from the holographic
p-wave insulator/superconductor models in the Yang-Mills theory, we
found in the AdS soliton background that the critical chemical
potentials are independent of the Weyl correction term, which tells
us that the correction to the Maxwell field will not affect the
properties of this new holographic p-wave insulator/superconductor
phase transition. We also observed that the effect of the Weyl
corrections cannot modify the critical phenomena, and found that
this new insulator/superconductor phase transition belongs to the
second order and the critical exponent of the system always takes
the mean-field value $1/2$. We confirmed our numerical result by
using the S-L analytic method and concluded that the Weyl
corrections have different effects on the holographic p-wave
insulator/superconductor phase transition of the Maxwell complex
vector field model and that of the Yang-Mills theory.

However, the story is completely different if we study the
holographic p-wave metal/superconductor phase transition with Weyl
corrections. We observed that similar to the effect of the Weyl
corrections in the Yang-Mills theory, in the black hole background,
the critical temperature for the vector operator increases as the
correction term increases, which implies that the higher Weyl
corrections make it easier for the vector condensation to form.
Further analytic studies showed that the holographic p-wave
metal/superconductor phase transition belongs to the second order
and the critical exponent of the system takes the mean-field value
$1/2$, which supports our numerical findings. Comparing the
holographic p-wave metal/superconductor phase transitions of the
Maxwell complex vector field model with that of the Yang-Mills
theory, we argued that these two p-wave models with Weyl corrections
share some similar features for the condensation of the vector
operator.

\begin{acknowledgments}

We thank Professor Elcio Abdalla for his helpful discussions and
suggestions. This work was supported by the National Natural Science
Foundation of China under Grant Nos. 11275066, 11175065 and
11475061; Hunan Provincial Natural Science Foundation of China under
Grant Nos. 12JJ4007 and 11JJ7001; and FAPESP No. 2013/26173-9.

\end{acknowledgments}

\end{document}